\documentclass[11pt]{article}
\usepackage{graphicx, caption, subfigure}
\makeatletter
\newcommand{\singlespacing}{\let\CS=\@currsize\renewcommand{\baselinestretch}{1.0}\tiny\CS}
\newcommand{\doublespacing}{\let\CS=\@currsize\renewcommand{\baselinestretch}{1.5}\tiny\CS}

\begin{document}
\title{On Production of Hadrons in Proton-Proton Collisions at RHIC and LHC Energies and an Approach }
\author{P.
Guptaroy$^1$\thanks{e-mail: gpradeepta@rediffmail.com}, Goutam
Sau$^2$\thanks{e-mail:  sau$\_$goutam@yahoo.com}  \&
S. Bhattacharyya$^3$\thanks{e-mail: bsubrata@isical.ac.in
}\\
{\small $^1$ Department of Physics, Raghunathpur College,}\\
 {\small P.O.: Raghunathpur 723133,  Dist.: Purulia(WB), India.}\\
{\small $^2$ Beramara Ram Chandrapur High School,}\\
 {\small South 24-Parganas, 743609(WB), India.}\\
 {\small $^3$ Physics
and Applied Mathematics Unit(PAMU),}\\
 {\small Indian Statistical Institute,}\\
 {\small 203 B. T. Road, Kolkata - 700108, India.}}
\date{}
\maketitle
\bigskip
\bigskip
\bigskip
\begin{abstract}
From the very early days of Particle Physics, both experimental and theoretical studies on proton-proton collisions
had occupied the center-stage of attention for very simple and obvious reasons. And this intense interest seems now
to be at peak value with the onset of the Large Hadron Collider (LHC)-studies at TeV ranges of energies.  In this work,
we have chosen to analyse the inclusive cross-sections, the rapidity density, the $K/\pi$ and $p/\pi$-ratio behaviours
and the $<p_T>$-values, in the light of the Sequential Chain Model (SCM). And the limited successes of the model
encourage us to take up further studies on several other aspects of topmost importance in particle physics with the same
approach.
\end{abstract}
\bigskip
\bigskip
\bigskip
 {\bf{Keywords}}: Relativistic heavy ion collisions, inclusive
production\\

 {\bf{PACS nos.}}: 25.75.q, 13.85.Ni
 \newpage
 \doublespacing
 \section{Introduction}
Proton-proton collisions are known to be the most elementary
interactions and form the very basis of our knowledge about the
nature of high energy collisions in general. Physicists, by and
large, hold the view quite firmly that the perturbative
quantum-chromodynamics (pQCD) provides a general framework for the
studies on high energy particle-particle collisions \cite{lappi}.
Obviously, the unprecedented high energies attained at Large Hadron
Collider (LHC) offer new windows and opportunities to test the
proposed QCD dynamics with its pros and cons. Naturally the normal
expectations run high that the bulk properties of the collision
system such as all the momentum spectra and correlations of all
produced hadrons should follow the strictures of QCD. But this not
definite and concretely-shaped knowledge about how this actually
happens and to what extent the process could be understood in the
perturbative and non-perturative domains. The issues involved here
still remain, to a considerable extent, quite open
\cite{schucraft},\cite{blibel}. Thus, having been somewhat repulsed
by the so-called standard approach, we try here to explain some
crucial aspects of measured data on $pp$ reactions at the LHC range
of energies with the help of some alternative approach. Our main
thrust would be on the properties of time-tested familiar
observables like transverse momenta spectra, rapidity distributions,
the ratio-behaviours and average transverse momenta ($<p_T>$) for
the charged secondaries in high energy $pp$ interactions. Comparison
with some other model would be made whenever possible.
\par
The organisation of the paper is as follows: In Section 2, we
provide a brief outline of the model chosen for study. In Section 3,
the results obtained by the model-based study are presented. In
Section 4, we end up with a discussion on the results and the
observations made in Section 3 and the conclusions.
 \section{The Approach: An Outline}
 This section gives a brief overview of the model-based features for
the production mechanism of the secondary hadrons in nucleon-nucleon
($p+p$) interaction in the context of the Sequential Chain Model
(SCM).
 According to
this Sequential Chain Model (SCM),  high energy hadronic
interactions boil down, essentially, to the pion-pion interactions;
as the protons are conceived in this model as
$p$~=~($\pi^+$$\pi^0$$\vartheta$), where $\vartheta$ is a spectator particle needed for
the dynamical generation of quantum numbers of the nucleons  \cite{pgr10}-\cite{bhat882}. The
production of pions in the present scheme occurs as follows: the
incident energetic $\pi$-mesons in the structure of the projectile
proton(nucleon) emits a rho($\varrho$)-meson in the interacting
field of the pion lying in the structure of the target proton, the
$\varrho$-meson then emits a $\pi$-meson and is changed into an
omega($\omega$)-meson, the $\omega$-meson then again emits a
$\pi$-meson and is transformed once again into a $\varrho$-meson and
thus the process of production of pion-secondaries continue in the
sequential chain of $\varrho$-$\omega$-$\pi$ mesons. The two ends of
the diagram contain the baryons exclusively \cite{pgr10}-\cite{bhat882}.
\par
For $K^+$($K^-$)or  $K^0{\bar{K^0}}$ production the model proposes
the following mechanism. One of the interacting $\pi$-mesons emits a
$\varrho$-mesons; the $\varrho$-mesons in its turn emits a
$\phi^0$-meson and a $\pi$-meson. The $\pi$-meson so produced then
again emits $\varrho$ and $\phi^0$ mesons and the process continues.
The $\phi^0$ mesons so produced now decays into either $K^+K^-$ or
$K^0{\bar{K^0}}$ pairs. The $\varrho$-$\pi$ chain proceeds in any
Fenymann diagram in a line with alternate positions, pushing the
$\phi^0$ mesons  (as producers of $K^+K^-$ or $K^0{\bar{K^0}}$
pairs) on the sides. This may appear paradoxical as the $\phi^0$
production cross-section is generally smaller than the $K{\bar K}$
production cross-section; still the situation arises due to the fact
that the $\phi^0$ resonances produced in the collision processes
will quickly decay into $K{\bar K}$ pairs, for which the number of
$\phi^0$ will be lower than that of the $K{\bar K}$ pairs. Besides,
as long as $\phi^0$ mesons remain in the virtual state,
theoretically there is no problem, for $\phi^0 K^+ K^-$ ( or $\phi^0
K^0 {\bar {K^0}}$) is an observed and allowed decay mode, wherein
the strangeness conservation is maintained with the
strange-antistrange coupled production. Moreover, $\phi^0 K^+ K^-$ (
or $\phi^0 K^0 {\bar {K^0}}$) coupling constant is well known and is
measured by experiments with a modest degree of reliability. And we
have made use of this measured coupling strength for our
calculational purposes, whenever necessary. It is assumed that the
$K^+K^-$ and $K^0{\bar{K^0}}$ pairs are produced in equal
proportions \cite{pgr10}-\cite{bhat882}. The entire production
process of kaon-antikaons is controlled jointly by the coupling
constants, involving $\varrho$-$\pi$-$\phi$ and $\phi^0$-$K^+K^-$ or
$\phi^0$-$K^0{\bar {K^0}}$.
\par
Now we describe here the baryon-antibaryon production. According to
the SCM mechanism, the decay of the pion secondaries produces
baryon-antibaryon pairs in a sequential chain as before. The pions
producing baryons-antibaryons pairs are obviously turned into the
virtual states. And the proton-antiproton pairs are just a part of
these secondary baryon-antibaryon pairs. In the case of
baryon-antibaryon pairs it is postulated that protons-antiprotons
and neutrons-antineutrons constitute the major bulk, Production of
the strange baryons-antibaryons are far less due to the much smaller values of the coupling constants and due to their being much heavier.
\par
The field theoretical calculations for the average multiplicities of
the $\pi$, $K$ and $\bar p$-secondaries and for the inclusive
cross-sections of those secondary particles deliver some expressions
which we would pick up from \cite{pgr10}-\cite{bhat882}.
\par
The inclusive cross-section of the $\pi^-$-meson produced in the
$p+p$ collisions given by
\begin{equation}\displaystyle E\frac{d^3\sigma}{dp^3}|_{pp \rightarrow
\pi^- x}  \cong \Gamma_{\pi^-} \exp(- 2.38 <n_{\pi^-}>_{pp}
x)\frac{1}{p_T^{(N_R^{\pi^-})}} \exp(\frac{-2.68
p_T^2}{<n_{\pi^-}>_{pp}(1-x)})   ~ ,
\end{equation}
with \begin{equation}\displaystyle {<n_{\pi^+}>_{pp} ~ \cong  ~
<n_{\pi^-}>_{pp} ~ \cong
 ~ <n_{\pi^0}>_{pp}  ~ \cong  ~ 1.1s^{1/5} ~,}
 \end{equation}
 where $\Gamma_{\pi^-}$ is the
normalisation factor which will increase as the inelastic
cross-section increases and it is different for different energy
region and for various collisions, for example, $|\Gamma_{\pi^-}| \cong
90$ for Intersecting Storage Ring(ISR) energy region. The terms
$p_T$, $x$ in equation (1) represent the transverse momentum,
Feynman Scaling variable respectively. Moreover, by definition, $x ~
= ~ 2p_L/{\sqrt s}$ where $p_L$ is the longitudinal momentum of the
particle. The $s$ in equation (2) is the square of the c.m. energy.
\par
$1/p_T^{N_R^{\pi^-}}$ of the expression (1) is the `constituent
rearrangement term' arising out of the partons inside the proton
which essentially provides a damping term in terms of a power-law in
$p_T$ with an exponent of varying values depending on both the
collision process and the specific $p_T$-range. The choice of
${N_R}$ would depend on the following factors: (i) the specificities
of the interacting projectile and target, (ii) the particularities
of the secondaries emitted from a specific hadronic or nuclear
interaction and (iii) the magnitudes of the momentum transfers and
of a phase factor (with a maximum value of unity) in the
rearrangement process in any collision. And this is a factor for
which we shall have to parameterize alongwith some physics-based
points indicated earlier. The parametrization is to be done for two
physical points, viz., the amount of momentum transfer and the
contributions from a phase factor arising out of the rearrangement
of the constituent partons. Collecting and combining all these, we
proposed the relation to be given by \cite{pgr08}
\begin{equation}\displaystyle
N_R=4<N_{part}>^{1/3}\theta,
\end{equation}
where $<N_{part}>$ denotes the average number of participating
nucleons and $\theta$ values are to be obtained phenomenologically
from the fits to the data-points. In this context, the only
additional physical information obtained from the observations made
here is: with increase in the peripherality of the collisions the
values of $\theta$ gradually grow less and less, and vise versa.
\par
Similarly, for kaons of any specific variety ( $K^+$, $K^-$, $K^0$
or $\bar{K^0}$ ) we have
\begin{equation}
\displaystyle E \frac{d^3\sigma}{dp^3}|_{pp \rightarrow K^- x}
 ~ \cong  ~ \Gamma_{K^-}\exp(  -  6.55 <n_{K^-}>_{pp}
 x) ~ \frac{1}{p_T^{(N_R^{K^-})}}\exp(\frac{-  1.33
  p_T^2}{<n_{K^-}>^{3/2}_{pp}}) ~ ~ ,
\end{equation}
 with $|\Gamma_{K^-}| \cong 11.22$ for ISR energies and with
\begin{equation}
\displaystyle <n_{K^+}>_{pp}   \cong  <n_{K^-}>_{pp}  \cong
  <n_{K^0}>_{pp}  \cong  <n_{\bar{K^0}}>_{pp} \cong
5\times10^{-2}  s^{1/4}  .
\end{equation}
And for the antiproton production in $pp$ scattering at high
energies, the derived expression for inclusive cross-section is
\begin{equation}
\displaystyle E\frac{d^3\sigma}{dp^3}|_{pp\rightarrow{\bar p}x}
 ~ \cong \Gamma_{\bar p} \exp(-25.4 <n_{\bar{p}}>_{pp} x)\frac{1}{p_T^{({N_R}^{\bar p})}}\exp(\frac{-0.66 ((p_T^2)_{\bar
p}+ {m_{\bar p}}^2)}{<n_{\bar p}>^{3/2}_{pp} (1-x)})
 ~  ,
\end{equation}
with $|\Gamma_{\bar p}| ~  \cong  ~ 1.87\times10^3$ and $m_{\bar p}$
is the mass of the antiprotons. For ultrahigh energies
\begin{equation}
\displaystyle{ <n_{\bar p}>_{pp}  ~ \cong <n_p>_{pp}  ~ \cong
 ~ 2\times10^{-2} ~ s^{1/4} ~ .}
 \end{equation}
\section{The Results}
Now let us proceed to apply the chosen model to interpret some recent experimental results of charged hadrons production for $p+p$ collisions at different energies. Here, the main observables are the inclusive cross-sections or invariant yields,
rapidity distributions, ratio behaviour and the average transverse momenta.
\subsection{Inclusive Cross-sections}
The general form of our SCM-based transverse-momentum distributions
for $p+p\rightarrow C^-+X$-type reactions can be written in the
following notation:
\begin{equation}\displaystyle{
E \frac{d^3\sigma}{dp^3}|_{pp \rightarrow C^- x}=\alpha_{C^-}\frac{1}{p_T^{N_R^{C^-}}}\exp(-\beta_{C^-} \times
p_T^2).}
\end{equation}
The value of $\alpha_{\pi^-}$, for example, can be calculated from the following relation:
\begin{equation}\displaystyle{
\alpha_{\pi^-}=\Gamma_{\pi^-}\exp(- 2.38
<n_{\pi^-}>_{pp} x)}
\end{equation}
The values of $(\alpha_{\pi^-})_{pp}$, $(N_R^{\pi^-})_{pp}$ and $(\beta_{\pi^-})_{pp}$  for different energies are given in Table 1. The experimental data for the inclusive cross-sections versus $p_T[GeV/c]$ for $\pi^-$ production in $p+p$
 interactions at $\sqrt{s_{NN}}$ = 62.4 GeV  and 200 GeV are taken
from Ref. \cite{phenix11} and they are plotted in Figs. 1(a) and 1(b) respectively. The production of $\pi^-$, $K^-$ and $\bar p$ at mid-rapidity in proton-proton
collisions at $\sqrt{s_{NN}}$ = 900 GeV  has been plotted  by lines in Fig. 1(c). Data are taken from  \cite{alice11}. For the data for charged particle distribution $Ed^3N_{ch}/dp^3=1/(2\pi p_T)E/p (d^2N_{ch}/d\eta dp_T)$ at energies $\sqrt{s_{NN}}$ = 546 GeV
and $\sqrt{s_{NN}}$ = 900 GeV we use
 references \cite{blibel}, \cite{ua5}. And for LHC data for charged particle distribution for energies $\sqrt{s_{NN}}$= 0.9 TeV, 2.36 Tev and 7 Tev we use references \cite{cms101}, \cite{cms102}. They are plotted
in Figure 2 and Figure 3 respectively. The solid
lines in those figures depict the SCM-based plots. As the main variety of the charged particles coming out are the pions, we use here eqn.(1) for calculational purposes. The `NSD'-term, used by the experimentalists, has the meaning of non-single diffractive collisions \cite{alice}.
\par
A comparison between the SCM-based results and the Tsallis parametrization is done for energies $\sqrt{s_{NN}}$= 0.9 TeV, 2.36 Tev and 7 Tev. The Tsallis parametric equation \cite{tsallis}, \cite{biro} is given hereunder
\begin{equation}\displaystyle{
E\frac{d^3N_{ch}}{dp^3} =\frac{1}{2\pi p_T}\frac{E}{p}\frac{d^2N_{ch}}{d\eta dp_T}=C\frac{dN_{ch}}{dy} (1+\frac{ET}{nT})^{-n},}
\end{equation}
with $y = 0.5 \ln[(E + p_z)/(E - p_z)]$, $E_T =\sqrt{m^2 + p^2_T}- m$ and $m$ is the charged pion mass. The dotted lines in Fig. 3 depict the Tsallis parametrization.
\par
Moreover, in Fig.4(a) and 4(b), we have plotted theoretical values $N_R$ and $\beta$ versus $\sqrt{s_{NN}}$ respectively.
\par
Similarly, by using eqn.(4), eqn.(5), eqn.(6) and eqn.(7), the values of $(\alpha_{K^-})_{pp}$, $(N_R^{K^-})_{pp}$, $(\beta_{K^-})_{pp}$ and $(\alpha_{\bar {p}})_{pp}$, $(N_R^{\bar {p}})_{pp}$, $(\beta_{\bar p})_{pp}$ are given in Table 2. The experimental data for the inclusive cross-sections versus $p_T[GeV/c]$ for $K^-$ and $\bar p$ production in $p+p$
 interactions at $\sqrt{s_{NN}}$ = 62.4 GeV ,200 GeV are taken
from Ref. \cite{phenix11} and for $\sqrt{s_{NN}}$ = 900 GeV we have used Ref. \cite{alice11}. They are plotted
in Figure 1(a), Figure 1(b) and Figure 1(c) respectively. The solid
lines in those figures depict the SCM-based plots.
\subsection{The Rapidity Distribution}
For the calculation of the rapidity distribution we can make use of a standard
relation as given below:
\begin{equation}\displaystyle{
\frac{dN}{dy}=\int (E \frac{d^3N_{ch}}{dp^3})dp_T }
\end{equation}
In Table 3 we had made a comparison between experimentally found $dn/dy$ for $\pi^-$, $K^-$ and $\bar p$ in $p+p$ collisions for RHIC  and LHC energies $\sqrt{s_{NN}}$ = 62.4 GeV, 200 GeV  and 900 GeV and the SCM-based calculated results. Data  are taken from refs. \cite{phenix11}, \cite{alice11} . The theoretically calculated results are coming out with the help of eqn. (1), eqn.(4), eqn.(6) and eqn. (11).
\par
Similarly, for LHC-energies, by using eqn.(1) and eqn. (11), the SCM-based $dN_{ch}/d\eta$ will be given hereunder
\begin{equation}\displaystyle{
\frac{dN_{ch}}{d\eta}=3.64 \exp(-0.007\sinh \eta)~~~~~for~~\sqrt{s_{NN}}~~ = ~~0.9~TeV,}
\end{equation}
\begin{equation}\displaystyle{
\frac{dN_{ch}}{d\eta}=4.75 \exp(-0.009\sinh \eta)~~~~~for~~\sqrt{s_{NN}}~~ = ~~2.38~TeV,}
\end{equation}
and
\begin{equation}\displaystyle{
\frac{dN_{ch}}{d\eta}=6.28 \exp(-0.011\sinh \eta)~~~~~for~~\sqrt{s_{NN}}~~ = ~~7~TeV.}
\end{equation}
In Fig. 5 we have plotted $dN_{ch}/d\eta$ vs. $\eta$ at three LHC-energies $\sqrt{s_{NN}}$ = 0.9 TeV, 2.38 TeV and 7 TeV. The reconstructed data points  for Fig. 5 are from Refs. \cite{cms101}, \cite{alice}, \cite{alice2}. Lines in the Figure are the outcomes of eqn. (12), eqn.(13) and eqn.(14) respectively.
\subsection{The Ratio-behaviours for Different Secondaries}
The nature of the relation of $K/\pi$ ratios with the SCM, presented in the previous work \cite{pgr02}, would be written in the following form
\begin{equation}\displaystyle{
\frac{K}{\pi}=5.4\times10^{-2}(\sqrt s)^{0.1}.}
\end{equation}
Fig. 6(a) shows the nature of rise of $K/\pi$ ratio in the light of SCM-based above relation (eqn. (15)). Data are taken from Ref. \cite{floris}.
\par
Similarly, in Fig. 6(b), we have presented the $p/\pi$ ratio for the RHIC and LHC-data.\cite{floris} The SCM-based calculations are done on the basis of eqn.(1) and Table 1.
\subsection{$<p_T>$ Values}
Next we attempt at deriving model-based expression for $<p_T>$.
\par
The definition for average transverse momentum  $<p_T>$ is
given below.
\begin{equation}\displaystyle{
<p_T>^C=\frac{\int^{p_T(max)}_{p_T(min)}p_TE{\frac{d^3\sigma}{dp^3}}^Cdp_T^2}{\int^{p_T(max)}_{p_T(min)}E{\frac{d^3\sigma}{dp^3}}^C
dp_T^2}, }
\end{equation}
The line in the Fig.7 depicts the SCM-based calculated results of the average transverse momentum
$<p_T>$ versus the c.m. energy $\sqrt{s_{NN}}$. The theoretical
calculations are done on the basis of uses of eqn.(1), Table 1. Data are taken from \cite{blibel}, \cite{cms101}, \cite{ua1}, \cite{e735}.
\section{Discussions and Conclusions}
Let us make some general observations and specific comments on the results arrived at and shown by the diagrams on a case-to-case basis.
\par
a) The measures of inclusive cross-sections against transverse momenta ($p_T$) obtained on the basis of the SCM for the pions, kaons and protons for the RHIC energies $\sqrt{s_{NN}}$ = 62.4, 200 GeV and for LHC energy $\sqrt{s_{NN}}$ = 900 GeV are depicted in Fig. 1. They describe a modest degree of success.
\par
Of the various types particles produced, the $\pi$-mesons, constitute, the near totality of the secondaries. So, in calculating the charged hadrons yields for different transverse momenta, on the basis of the model, for different energies ranging from $\sqrt{s_{NN}}$ = 546 GeV  to 7 TeV, we use eqn. (1). The results are shown in the Fig. 2 and Fig. 3. Moreover, we have compared the Model-based results with Tsallis parametrization in Fig.3. The outcomes of these plots are fairly satisfactory.
\par
b) The theoretically calculated $N_R$ and $\beta$-factors of eqn. (8) for different $\sqrt{s_{NN}}$'s have been plotted in Fig.4. With the inclusion of the power law form arising due to the physics of partonic rearrangement factor, the model has turned effectively into a mixed model. And these two factors corroborate effectively the ``soft" and ``hard" regimes without any extra effort.
\par
c) The calculations of rapidity distributions for pions, kaons and
protons at different RHIC  and LHC energies, on the basis of the
model, have been done with the help of eqn. (11). The calculated
values are compared with the experimental ones and they are shown in
Table 3. The theoretical values are in fair agreement with the
experimental data.
\par
Similarly, in Fig. 5, we have plotted the pseudorapidity distributions for different LHC energies.
Here, the model modestly reproduce data consistently.
\par
d) The agreements between the measured data on $K/\pi$ and $p/\pi$
ratio and the theoretical SCM plots (shown in Fig. 6)for different
energies are strikingly encouraging.
\par
e) Fig. 7 shows the plots of $<p_T>$ vs. $\sqrt{s_{NN}}$. The initial indication of the SCM-based theoretical plot
shows a modest agreement with the data.
\par
Finally, we conclude from the analysis of the results given above
with the following statements: The model applied here gives fair
descriptions of the $p_T$-spectra of all the light secondaries or
charged hadron. Some disagreements are observed at pseudorapidity
distributions at LHC energies. The model might require some finer
adjustments to cope with the data for very high energy nuclear
collisions.  However, on an overall basis, our model is in fair
agreement with the latest $pp$-collision results obtained from the
uptodate LHC experiments. This factor is really of high important to
us. Furthermore, the observables or the physical aspects that we
have reckoned herewith form a clear continuum from the old ISR
experiments to the recent Large Hadron Collider studies via the
intermediary BNL-RHIC results cropped up over the 1st decade of this
century. In so far as the rolls of the other models (including QCD
versions) are considered, the results do neither speak very high
about them; rather they cast doubts on the suitability of them in
applying at this LHC energy band. And this certainly spurs us on to
take up further studies on the SCM proposed earlier and applied in
the present study.
\newpage
\singlespacing

\newpage
\begin{table}
\caption{Values of $(\alpha_{\pi^-})_{pp}$, $(N_R^{\pi^-})_{pp}$ and $(\beta_{\pi^-})_{pp}$ for
$\pi^-$ productions in $p+p$ collisions at
$\sqrt{s_{NN}}$=62.4, 200, 546, 900, 2380 and 7000 GeV}
\begin{center}
\begin{tabular}{cccc}
\hline
$\sqrt{s_{NN}}$&$(\alpha_{\pi^-})_{pp}$&($N_R^{\pi^-})_{pp}$&$(\beta_{\pi^-})_{pp}$\\
\hline
62.4 GeV& 0.545&3.327&0.468\\
\hline
200 GeV& 0.907&3.867&0.293\\
\hline
546 GeV&0.153&3.931&0.172\\
\hline
900 GeV&0.135&4.154&0.128\\
\hline
2380 GeV&0.166&4.235&0.085\\
\hline
7000 GeV&0.355&4.366&0.075\\
\hline
\end{tabular}
\end{center}
\end{table}

\begin{table}
\caption{Values of $(\alpha_{K^-})_{pp}$, $(N_R^{K^-})_{pp}$, $(\beta_{K^-})_{pp}$  and $(\alpha_{\bar p})_{pp}$, $(N_R^{\bar p})_{pp}$ and $(\beta_{\bar p})_{pp}$ for
$K^-$, $\bar p$ productions in $p+p$ collisions at
$\sqrt{s_{NN}}$=62.4, 200 GeV and 900 GeV}
\begin{center}
\begin{tabular}{cccc}
\hline
$\sqrt{s_{NN}}$&$(\alpha_{K^-})_{pp}$&($N_R^{K^-})_{pp}$&$(\beta_{K^-})_{pp}$\\
\hline
62.4 GeV& 0.245&2.527&0.591\\
\hline
200 GeV& 0.235&3.017&0.417\\
\hline
900 GeV& 0.047&1.544&0.248\\
\hline
\hline
$\sqrt{s_{NN}}$&$(\alpha_{\bar p})_{pp}$&($N_R^{\bar p})_{pp}$&$(\beta_{\bar p})_{pp}$\\
\hline
62.4 GeV&0.215&1.527&0.618\\
\hline
200 GeV&0.115&2.117&0.426\\
\hline
900 GeV& 0.027&1.344&0.248\\
\hline
\end{tabular}
\end{center}
\end{table}

\begin{table}
\caption{Comparisons of experimental $dn/dy$ with the SCM-based theoretical ones for $\pi^-$,
$K^-$, $\bar p$ productions in $p+p$ collisions at
$\sqrt{s_{NN}}$=62.4 GeV, 200 GeV and 900 GeV}
\begin{center}
\begin{tabular}{cccc}
\hline
$\sqrt{s_{NN}}$& particle &$dn/dy$&$dn/dy$\\
&&(Experimental)&(Theoretical)\\
\hline
62.4 GeV& $\pi^-$&0.900$\pm$0.063&0.840\\
&$K^-$&0.103$\pm$0.005&0.100\\
&$\bar p$&0.037$\pm$0.003&0.035\\
\hline
200 GeV& $\pi^-$&0.824$\pm$0.053&0.841\\
&$K^-$&0.067$\pm$0.003&0.065\\
&$\bar p$&0.022$\pm$0.002&0.018\\
\hline
900 GeV& $\pi^-$&1.485 $\pm$ 0.004&1.241\\
&$K^-$&0.182 $\pm$ 0.004&0.175\\
&$\bar p$&0.079 $\pm$ 0.002&0.068\\
\hline
\end{tabular}
\end{center}
\end{table}
\newpage
\begin{figure}
\subfigure[]{
\begin{minipage}{.5\textwidth}
\centering
\includegraphics[width=2.5in]{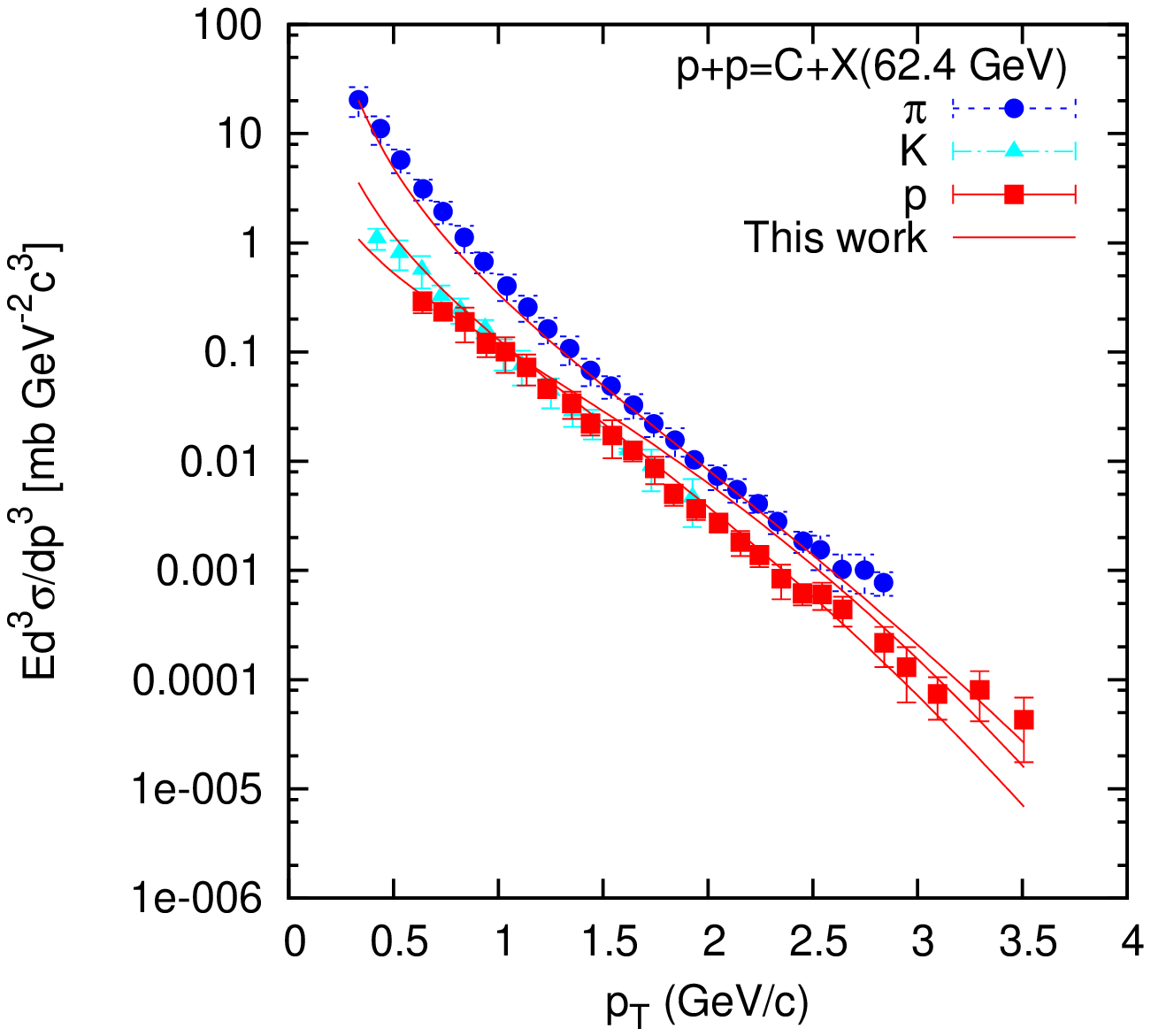}
\setcaptionwidth{2.6in}
\end{minipage}}%
\subfigure[]{
\begin{minipage}{0.5\textwidth}
\centering
 \includegraphics[width=2.5in]{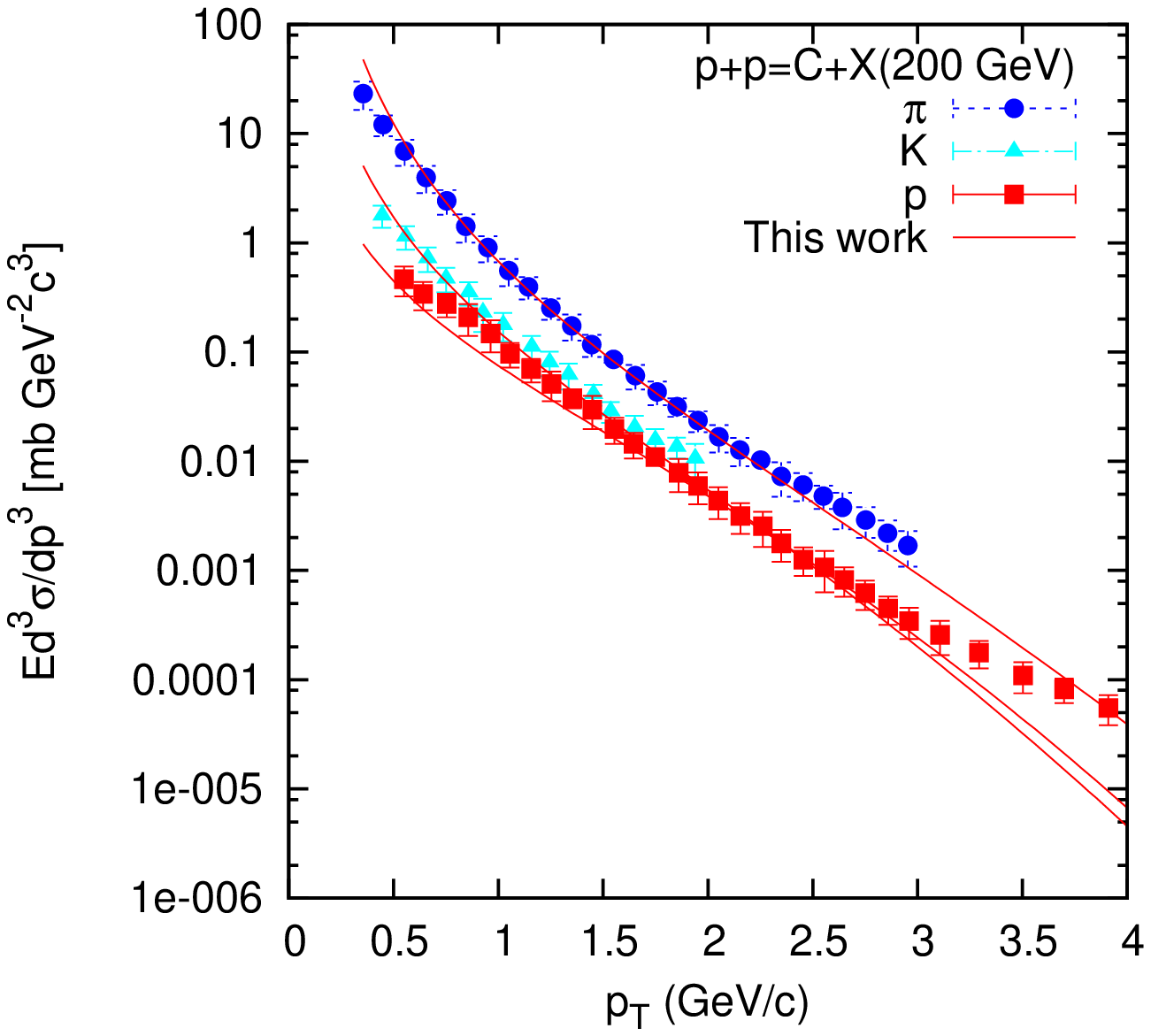}
  \end{minipage}}
  \vspace{0.01in}
 \subfigure[]{
\centering
 \includegraphics[width=2.5in]{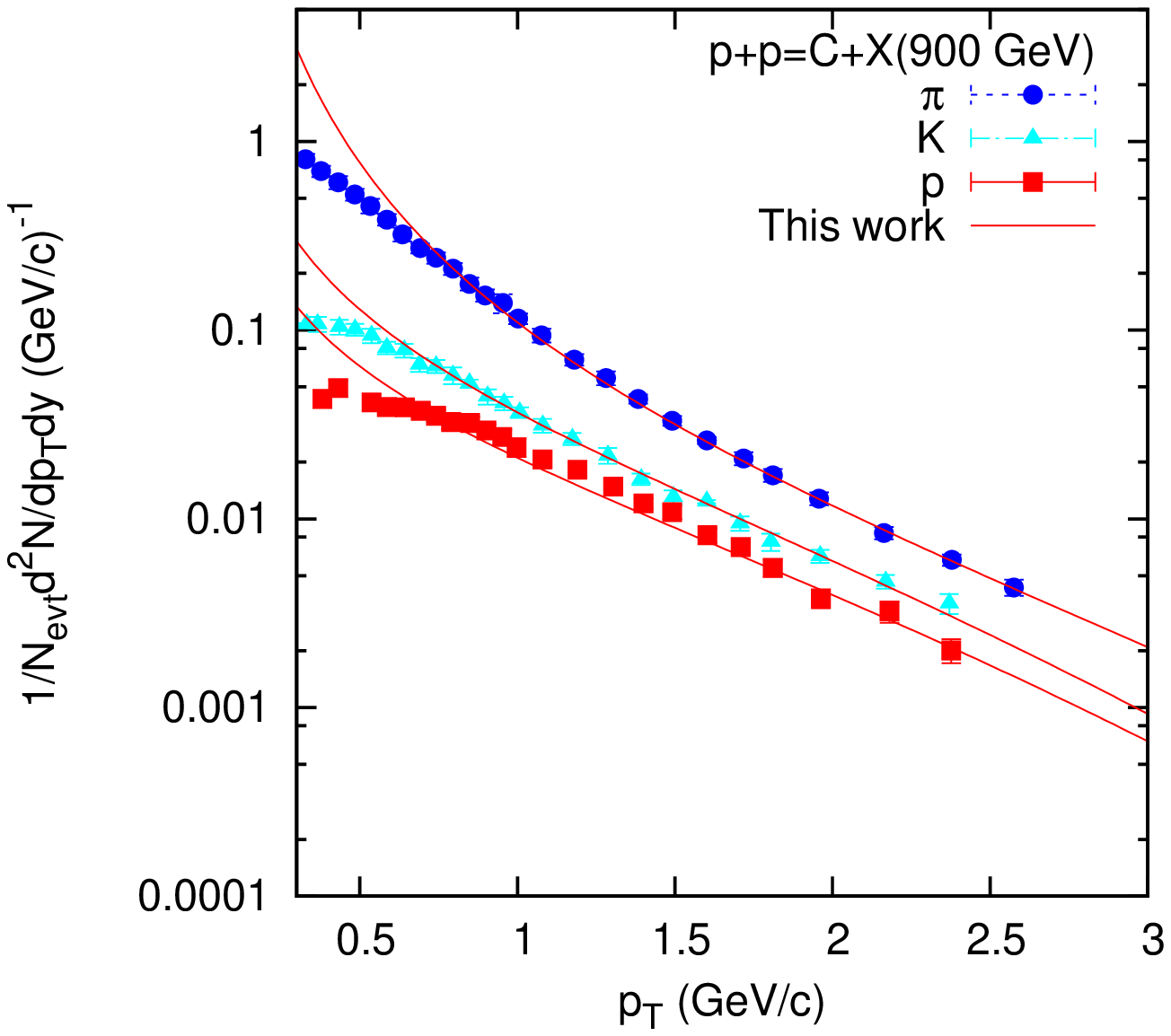}}
  \caption{\small Plots for $\pi$, $K$ and proton-production in $p+p$ collisions at RHIC
energies (a) $\sqrt{s_{NN}}$ = 62.4 GeV, (b)$\sqrt{s_{NN}}$ = 200 GeV and (c) $\sqrt{s_{NN}}$ = 900 GeV. Data are taken from Ref.
\cite{phenix11} for Figs. (a) and (b), while for Fig. (c) from Ref. \cite{alice11}. Solid lines in the Figures show the SCM-based
 plots.   }
\end{figure}
\begin{figure}
\begin{minipage}{.5\textwidth}
\includegraphics[width=3in]{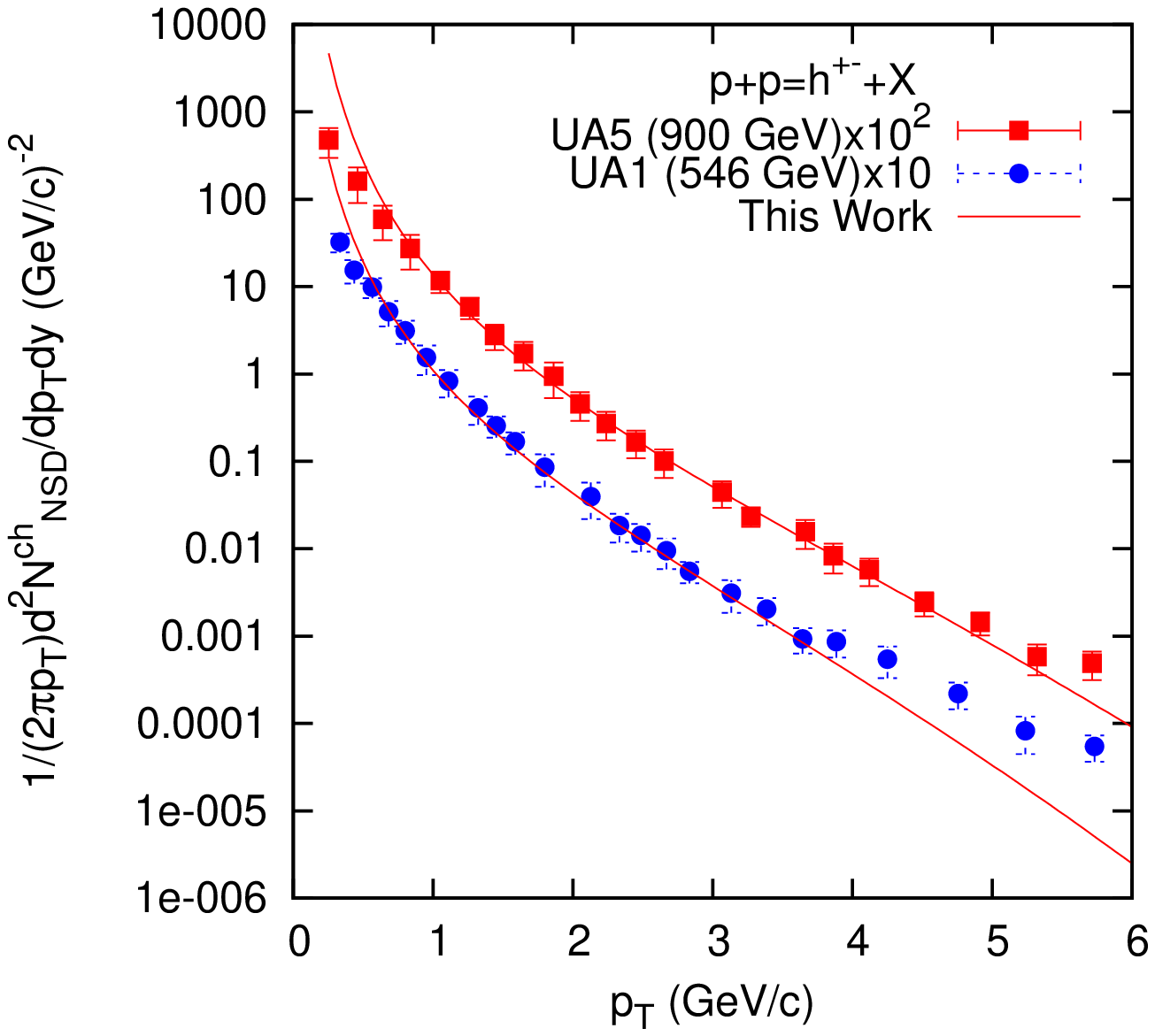}
\setcaptionwidth{2.8in}\caption{\small Transverse momentum distributions
of the invariant cross section of charged particles in NSD $p+p$
collisions at $|y|\leq 2.5$ for energies $\sqrt{s_{NN}}$ = 546 GeV and 900 GeV. Experimental data are taken from \cite{blibel}, \cite{ua5}. Lines show the theoretical plots. }
\end{minipage}%
\begin{minipage}{.5\textwidth}
  \centering
\includegraphics[width=3in]{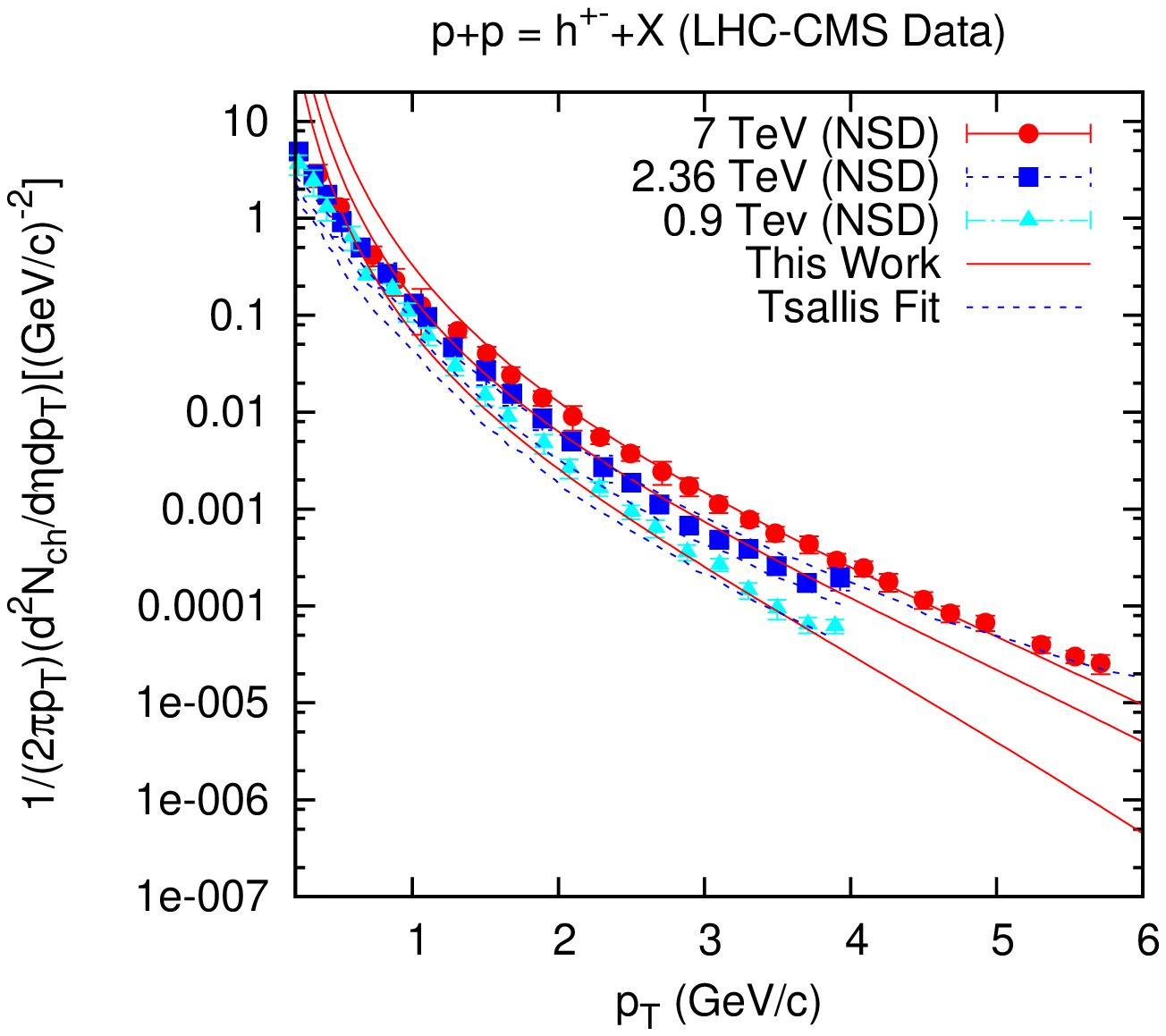}
\setcaptionwidth{2.8in} \caption{\small Charged-hadron yield for energies $\sqrt{s_{NN}}$ = 0.9 TeV , 2.38 TeV and 7 TeV in the range $\eta < 2.4$ in NSD events as a function of $p_T$; Data are taken from CMS collaboration \cite{cms101}-\cite{cms102}. Solid lines in the Fig. represent SCM-based results while the dashed lines show Tsallis fit.  }
\end{minipage}%
\end{figure}
\begin{figure}
\subfigure[]{
\begin{minipage}{.5\textwidth}
\centering
\includegraphics[width=2.5in]{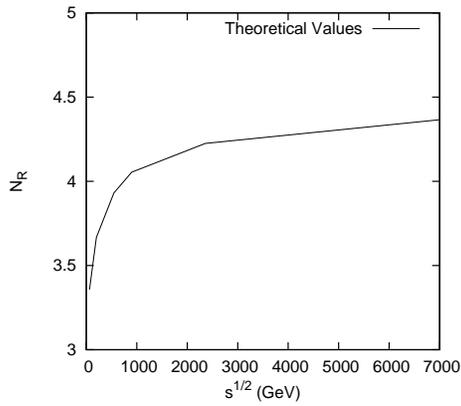}
\setcaptionwidth{2.6in}
\end{minipage}}%
\subfigure[]{
\begin{minipage}{0.5\textwidth}
\centering
 \includegraphics[width=2.5in]{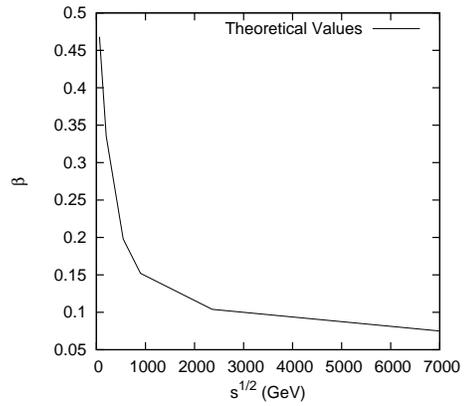}
  \end{minipage}}
  \caption{\small Plots of theoretical (a) $N_R$ and (b) $\beta$-values versus $\sqrt s$.  }
\end{figure}
\begin{figure}
\centering
\subfigure{
\begin{minipage}{.244\textwidth}
\centering
\includegraphics[width=1.8in]{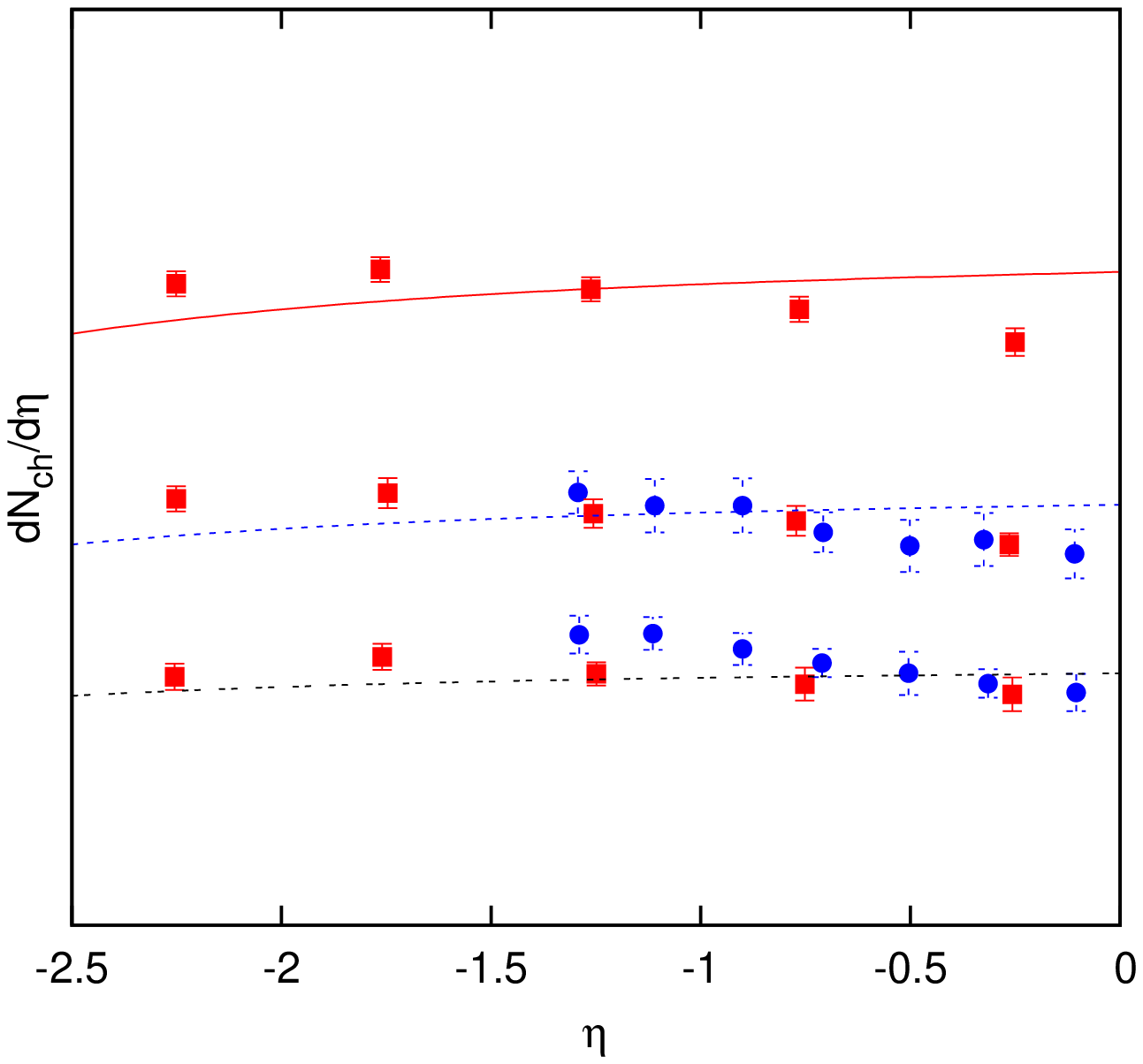}
\end{minipage}}%
\subfigure{
\begin{minipage}{0.24\textwidth}
\centering
 \includegraphics[width=1.8in]{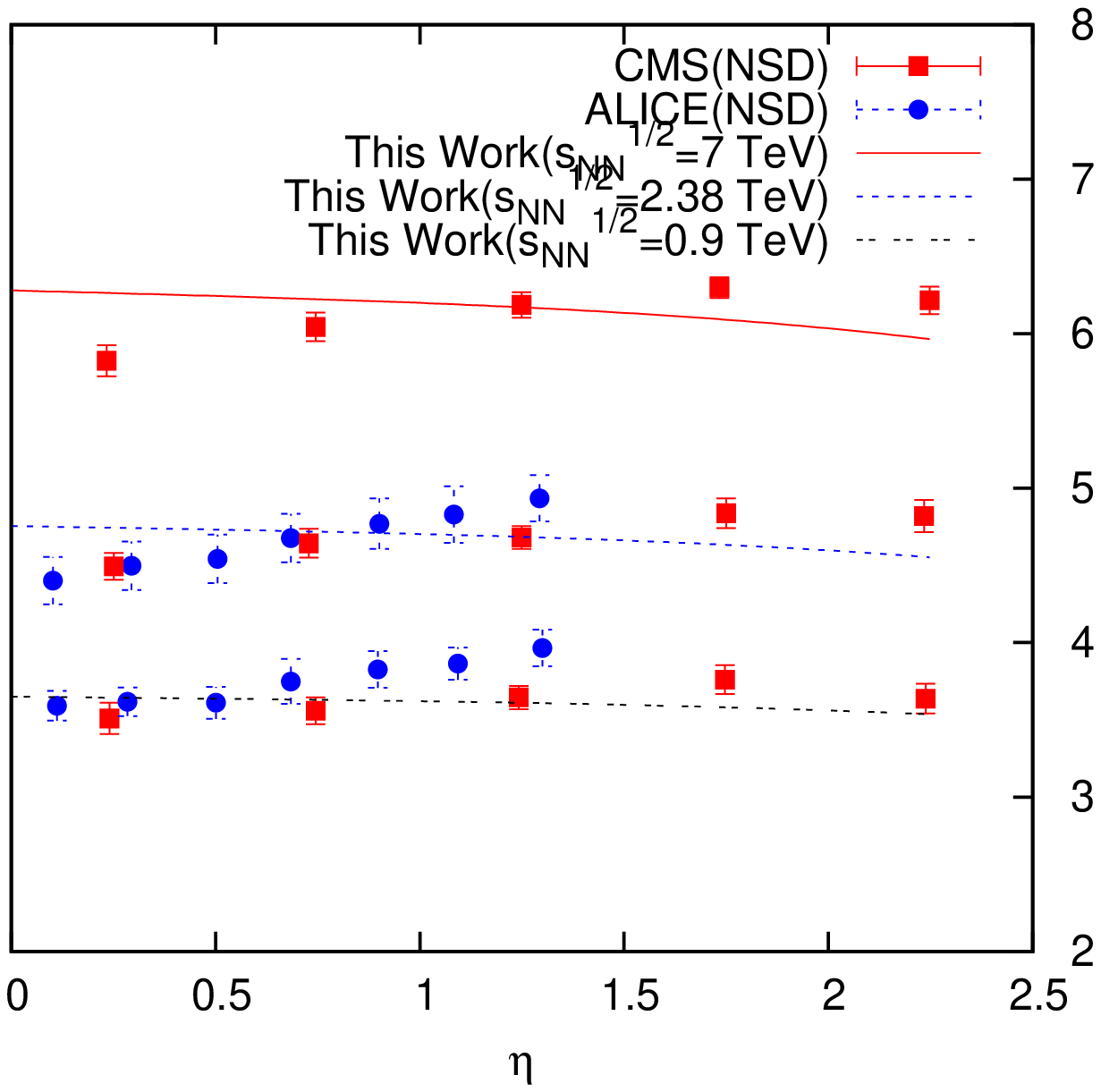}
  \end{minipage}}%
   \caption{\small Plot of $dN_{ch}/d\eta$ vs. $\eta$ for $p+p$ collisions at $\sqrt{s_{NN}}$ = 0.9 TeV, 2.38 TeV and 7 TeV. The reconstructed data points are from Refs. \cite{cms101}, \cite{alice},\cite{alice2}. The lines in the figure depict the theoretical results for different energies. }
\end{figure}
\begin{figure}
\subfigure[]{
\begin{minipage}{.5\textwidth}
\centering
\includegraphics[width=2.5in]{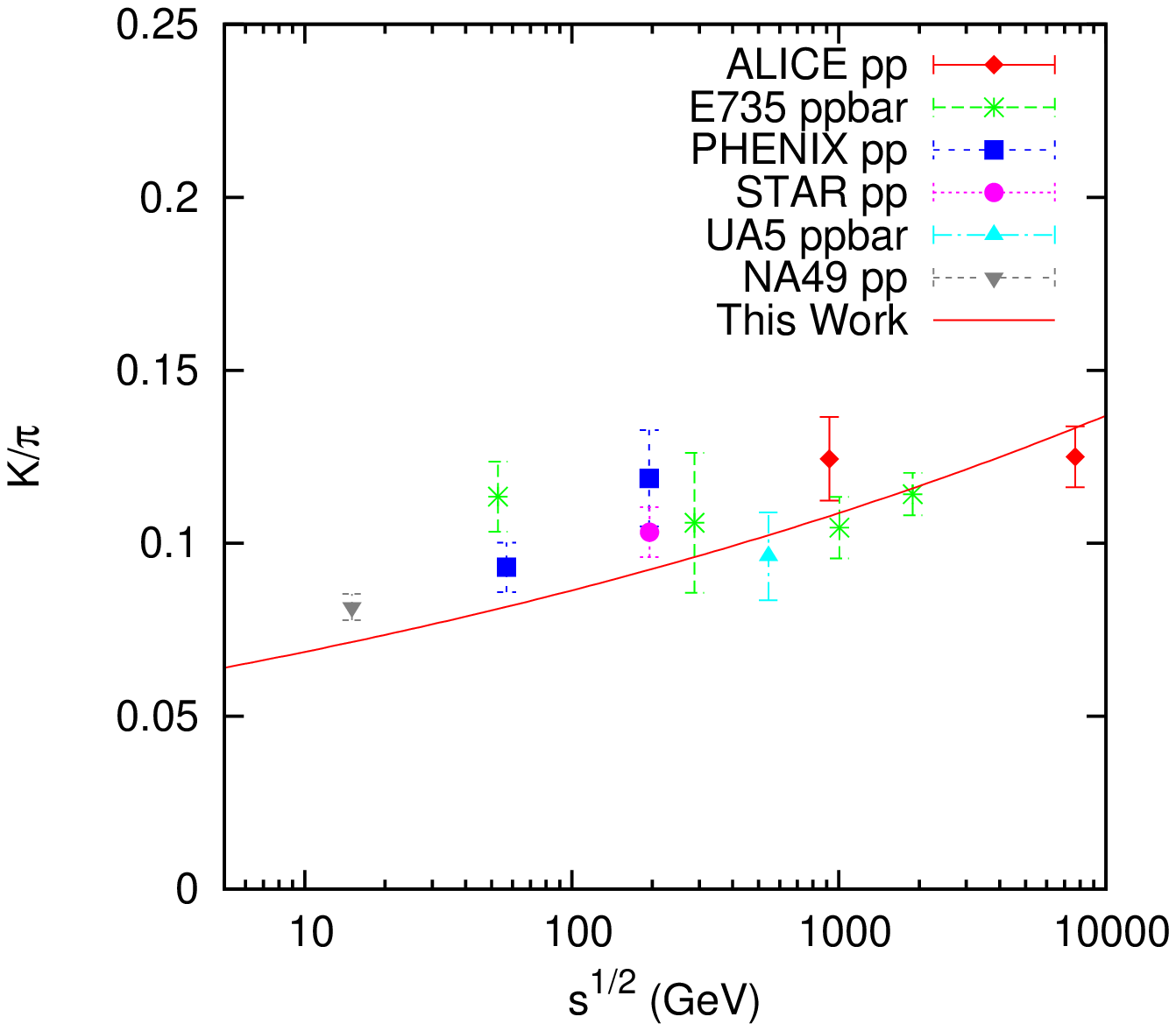}
\setcaptionwidth{2.6in}
\end{minipage}}%
\subfigure[]{
\begin{minipage}{0.5\textwidth}
\centering
 \includegraphics[width=2.5in]{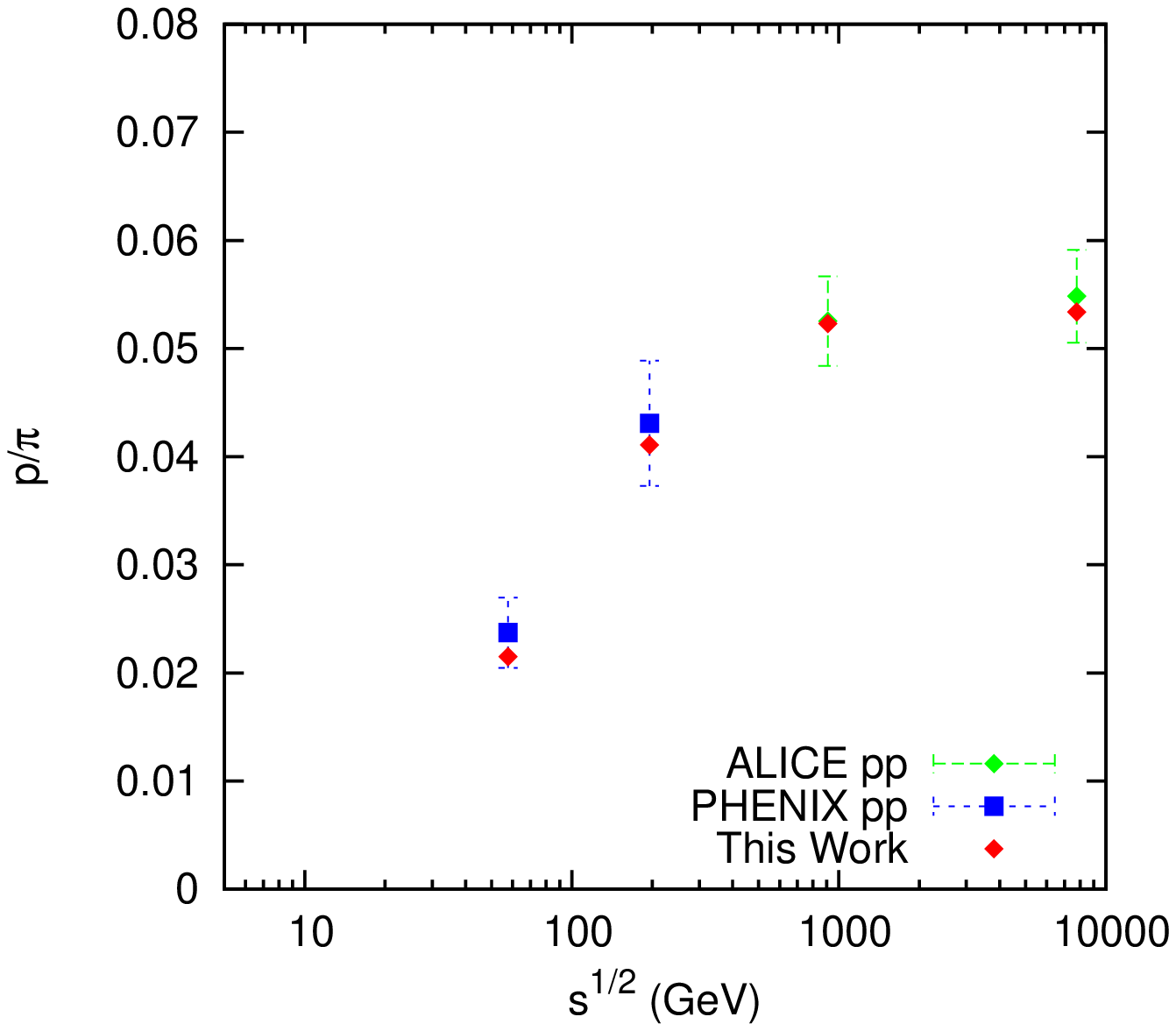}
  \end{minipage}}
  \caption{\small  Presentation of plots of (a) $K/\pi$ and (b) $p/\pi$ at different center-of-mass energies. Line and rhombus represent the SCM-based results in (a) and (b) respectively against the data sets taken from Ref. \cite{floris} }
\end{figure}
\begin{figure}
\centering
\includegraphics[width=2.5in]{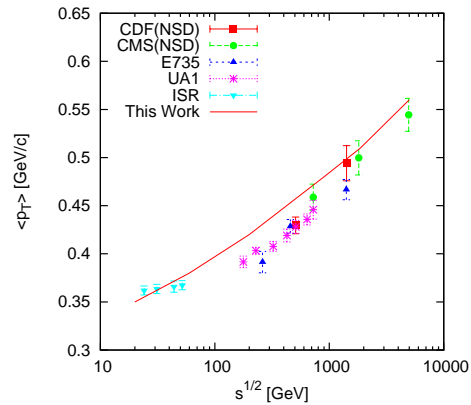}
  \caption{\small  Plot of average transverse momentum as function of  $\sqrt{s}$. Data are taken from \cite{blibel}, \cite{cms101}, \cite{ua1}, \cite{e735}. Line shows the SCM-based calculations. }
\end{figure}
 \end{document}